# Electron emission from conduction band of diamond with negative electron affinity

H. Yamaguchi[1,2], T. Masuzawa[3], S. Nozue[3], Y. Kudo[3], I. Saito[4], J. Koe[3], M. Kudo[5], T. Yamada[6], Y. Takakuwa[7], and K. Okano[2,3]

[1] *Department of Materials Science and Engineering, Rutgers University, 607 Taylor Road, Piscataway, NJ 08854, USA*

[2] *School of Materials Science, Japan Advanced Institute of Science and Technology, 1-1, Asahidai, Nomi, Ishikawa 923-1292, Japan*

[3] *Department of Material Science, International Christian University, 3-10-2 Osawa, Mitaka, Tokyo 181-8585, Japan*

[4] *Department of Engineering, University of Cambridge, 9, JJ Thomson Avenue, Cambridge CB3 0FA, U.K.*

[5] *JEOL Ltd., 3-1-2 Musashino, Akishima, Tokyo 196-8558, Japan*

[6] *Diamond Research Centre, National Institute of Advanced Industrial Science and Technology (AIST), 1-1-1 Umezono, Tsukuba, Ibaraki 305-8568, Japan*

[7] *Division of Materials Control, Institute of Multidisciplinary Research for Advanced Materials, Tohoku University, 2-1-1 Katahira, Aoba, Sendai 980-8577, Japan*

*Corresponding author: kenokano@icu.ac.jp*

## Abstract

Experimental evidence explaining the extremely low-threshold electron emission from diamond reported in 1996 has been obtained for the first time. Direct observation using combined ultraviolet photoelectron spectroscopy/field emission spectroscopy (UPS/FES) proved that the origin of field-induced electron emission from heavily nitrogen (N)-doped chemical vapour deposited (CVD) diamond was at conduction band minimum (CBM) utilising negative electron affinity (NEA). The significance of the result is that not only does it prove the utilisation of NEA as the dominant factor for the extremely low-threshold electron emission from heavily N-doped CVD diamond, but also strongly implies that such low-threshold emission is possible from other types of diamond, and even other materials having NEA surface. The low-threshold voltage, along with the stable intensity and remarkably narrow energy width, suggests that this type of electron emission can be applied to develop a next generation vacuum nano-electronic devices with long lifetime and high energy resolution.

## Introduction

Electron field emission from a diamond surface has attracted much attention, due to the fact that this material was reported to exhibit negative electron affinity (NEA) [1-3]. Furthermore, the development of the chemical vapour deposition (CVD) technique for growing diamond at atmospheric pressure [4] accelerated research in this field [5-6]. Although the threshold voltage/field was not as low as expected from theory, there have been many reports concerning electron emission from this surface [7-13], largely because the diamond surface is extremely chemically inert. The inertness prevents the diamond surface from aging, a process often observed in field emission studies of metals [14]. Attempts to utilise NEA did not quite lead to the fabrication of practical cathodes made of diamond, due to difficulties in achieving the n-type electrical conduction, which seemed to be the only solution to realise the predicted low-threshold electron emission from the NEA surface.

Reports concerning electron emission from diamond can be categorised into three main types from the perspective of the emitted electrons' origin. These are:

(i) emission from the conduction band (CB) utilising NEA

(ii) emission from the impurity band (IB)

(iii) emission from the valence band (VB)

The expected scenario for achieving low-threshold emission is only possible in Type (i) because the electron affinity plays a less important role in Type (ii) and makes almost no contribution to the electron emission categorised in Type (iii). The reason is that the energy origin of the emitted electrons is too low and the barrier formed at the diamond/vacuum interface generally prevents low-threshold emission.

There are only a few reports, which might be categorised into Type (i). The first was electron emission from heavily N-doped CVD diamond reported by one of the present authors (KO) in 1996 [15]. The second is the triple junction emission reported by Geis in 1998 [16]. Theoretical interpretations of these experimental results were made by many researchers, reflecting the importance of these two reports [17-20]. To the best of the present authors' knowledge, all the other reports describe emission at higher threshold voltage/field, which implies that they can be categorised as Type (ii) or (iii). Type (ii) consists of CVD diamonds doped with impurities. Although the origin differs slightly, emission from the impurity band was reported from P- and lightly N-doped CVD diamonds [21-22]. A typical example which was shown to be Type (iii) was the natural IIb diamond. The origin was found to be at the VBM by Bandis and Pate in 1996 [23]. The majority of reports on field emission from diamond describe Type (ii), or (iii), and these discouraged extensive experimental research trying to use diamonds as a cathode in next generation vacuum nano-eletronic devices.

This report presents experimental evidence for the origin of the field-induced electron emission from heavily N-doped CVD diamond, which proves that the emission can be categorised as Type (i). The results clearly describe the unique origin of the "extremely low-threshold electron emission," for which the phenomenon itself was reported in 1996 [15]. In addition, the results showed the electrons emitted from the diamond exhibit stable intensity and remarkably narrow energy width, two factors vital for next generation vacuum nano-eletronic devices with long lifetime and excellent high-energy resolution performance.

**Experimental**

In order to clarify the origin of emitted electrons, X-ray photoelectron spectroscopy (XPS), ultraviolet photoelectron spectroscopy (UPS) and field emission spectroscopy (FES) were combined and simultaneously performed. The significance of combined XPS/UPS/FES is that the origin of the emitted electrons can directly be defined with high spatial and energy resolution. The origin can be defined with an energy resolution in the order of ~0.1 eV. The schematic diagram of the system is shown in Figure 1. Mg K$\alpha$ (h$\upsilon$=1253.6 eV), and He I (h$\upsilon$=21.2 eV) were used as X-ray and UV light sources, respectively. An electrically grounded nickel (Ni) mesh grid was placed 100 μm above the diamond surface to apply a negative voltage for FES.

Measurements were performed in an ultra high vacuum (UHV) at a base pressure of 5.0 x $10^{-10}$ Torr. A hemispherical analyser with a radius of 100 mm was used for analysing the kinetic energy of emitted electrons. The energy resolution was 0.15 eV from the width of the Fermi edge obtained from a gold (Au) reference sample at room temperature.

Heavily N-doped polycrystalline diamond was grown on a Si substrate for two hours by the hot filament chemical vapour deposition (HFCVD) technique. The growth conditions were as follows: the filament temperature was 2300℃, substrate temperature was 850℃, reaction pressure was 100 Torr, and the ratio of reactant gas to hydrogen was 0.6 vol.%. A saturated solution of urea and methanol was diluted with acetone, and vaporised to be used as a reactant gas to achieve a nitrogen concentration of ~$10^{20}$cm$^{-3}$ [15].

## Results and Discussion

The combined UPS/FES approach used in this study consisted of two techniques to induce electron emission. One was the application of UV light to induce photoemission for UPS, and the other was the application of an electric field to induce electron emission for FES. The uniqueness of the combined UPS/FES is that the spectrum, which contains the energy band structure of diamond and the energy of the emitted electrons, can be obtained under the same experimental conditions. The spectrum evidences the origin by comparing the energy band structure and the energy of the emitted electrons, free from other effects, including the photovoltaic effect, which generally prevents precise energy alignment [24].

In order to perform combined UPS/FES, a voltage was initially applied between the mesh grid and the diamond to induce electron emission for FES. The voltage was increased from 0 V at approximately 50 V steps, and field-induced emission (FE) spectra were taken at each applied voltage. A FE peak appeared at 307 V, and its intensity and kinetic energy increased corresponding to the voltage. The width of peak from site 1 was as narrow as ~0.3 eV, one of the narrowest energy widths available for an electron beam at room temperature. The appearance of a single peak (from site 1) was consistent with our previous results on natural IIb diamond and lightly N-doped CVD diamond, as illustrated in Figure 2. A single peak was observed for each type of the diamond, while their origins were different [25-26]. Unique to heavily N-doped CVD diamond was the appearance of the second FE peak (from site 2), which appeared at an applied voltage of 512 V and above. The second peak also increased its intensity and shifted to higher energy corresponding to the voltage. However, the energy difference between the two peaks (from site 1 and 2) increased as applied voltage increased. A detailed analysis regarding the change in the difference will be described later. The tail, which cannot be observed on the lower side because of the sharp cutoff due to CBM, reflects the broader electron distribution above the CBM. Moreover, the results suggest that there might be a scattering process in the field-induced electron emission. The rather broad width for peak from site 2 can often be observed as the result of the scattering process. The reason of existing two peaks was speculated to be due to the difference in surface electronic conditions [27-32]. The detailed analysis is to be performed by adding XPS to the UPS/FES, and will be reported elsewhere.

After FES was successfully performed, the sample was irradiated with UV light to measure combined UPS/FES, finally defining the origin of field-induced electron emission from heavily N-doped CVD diamond. As soon as the diamond surface was illuminated by UV light, it glowed blue-violet due to the light emitted by the recombination of photo-excited electron and hole pairs. In the UPS/FES spectrum, broad double peaks appeared with the characteristic shape for diamond, in addition to the FE peaks. Figure 3 shows the result, which was obtained for voltages between 512 and 716 V. The abscissa shows the kinetic energy of the emitted electrons relative to the VBM in eV, and the ordinate shows the intensity in arbitrary units. The most important result in the combined UPS/FES is the agreement between FE peak from site 2 of the FES and the low-energy cutoff of the UPS peak, which is clear even without detailed analysis of the spectra. The match clearly shows that the origin of the emitted electrons was at the CBM. Furthermore, analysis of the UPS peak indicates that the electron affinity (EA) of the diamond was negative, with a value of EA=~-0.8 eV, comparable to that of literature [3, 28]. Hence, the combined UPS/FES proved that the electrons were emitted from the CB utilising NEA, and that the electron emission from heavily N-doped CVD diamond can be categorised as Type (i).

As the applied voltage was increased, the intensity of peak from site 2 increased similar to that of the individual FES. However, the correlation was consistently observed, regardless of the applied voltage, which suggests a continuous supply of sufficient electrons in the CB during field-induced electron emission. As mentioned earlier, the slight broadening of peak from site 2 suggests a scattering process, which may be explained by the metal-insulator-vacuum (MIV)-type electron emission proposed using the emission current – anode voltage (I-V) characteristics of heavily N-doped CVD diamond [19]. In MIV-type electron emission, the voltage applied between an anode and a cathode drops in diamond bulk, enhancing electron injection from the back contact metal to the CB of diamond due to quantum tunnelling. These electrons move through the diamond bulk to the surface, where they are emitted into the vacuum. It is possible that scattering during electron transport through the diamond bulk may be responsible for the observed peak broadening. MIV-type electron emission also explains the role of voltage required to induce electron emission from heavily N-doped CVD diamond. The voltage is necessary to induce an electric field in the diamond bulk in order to enhance electron injection from the back contact metal to the CB of diamond, although there is no energy barrier at the diamond/vacuum interface. The applied voltage of a few hundred volts was consistent with the voltage needed to induce a sufficient electric field in diamond bulk. The shift observed for the first peak (from site 1) is speculated to be due to the voltage drop induced by the emission current. When the peak energy is plotted against emission current, the data fits to a straight line, where the slope indicates the resistance of the diamond bulk. In order to further investigate the observed shift, the emission current dependence of the FE peak for diamonds with different resistivity was compared.

Figure 4 shows the emission current dependence of FE peak energy for heavily N-doped CVD diamond, lightly N-doped CVD diamond, and natural IIb diamond. Plots for each FE peak can be fitted to a straight line, where the slopes indicate the resistance of the diamond bulk. The slope indicated resistances of $\sim 10^{11}\ \Omega$ and $\sim 10^{9}\ \Omega$ for heavily and lightly N-doped CVD diamond, respectively, which is consistent with the resistivity obtained using the conventional four-point probe method [19]. The slope was almost zero for natural IIb diamond because the voltage drop was less than the energy resolution of the measurement system. This was also consistent with the small voltage drop expected for natural IIb diamond with resistivity of $\sim 10^{2}\ \Omega$ cm.

After confirming that the plot slopes correlate to the resistivity of each diamond,

we applied the energy shift observed for the FE peak to determine the "true" origin of the emitted electrons. Although the combined UPS/FES spectra for each applied voltage should indicate the origin of the emitted electrons, the voltage dependence of the FE peaks clearly suggests that they were affected by the voltage drop induced by the emission current. In order to define the "true" origin of the electrons emitted from the diamond, the origin without the effect of the voltage drop must be explored. For this purpose, we extrapolated the FE peak energy - emission current characteristics to "emission current=0." The energy level for "emission current=0" provides the "true" origin free from the voltage drop induced by the emission current. It must be stressed here that this extrapolation method is a method, which permits definition of the "true" origin because from the experimental point of view, the signal intensity for combined UPS/FES cannot be obtained for “emission current=0”. The "true" origin determined using the extrapolation method clearly indicated that electrons are emitted from the CBM utilising NEA for heavily N-doped CVD diamond. Both peaks from site 1 and 2 can be extrapolated to the CBM, which is ~5.5 eV above the VBM.

Our recent computer simulation shows that a nitrogen concentration of ~$10^{20}cm^{-3}$ is required to achieve an energy barrier of as narrow as 0.7nm at the diamond - metal interface [33]. The barrier width of 0.7nm is narrow enough to induce quantum electron tunnelling for sufficient emission current, and the nitrogen concentration of ~$10^{20}cm^{-3}$ is possible with heavily N-doped CVD diamond. Therefore sufficient electron injection at the diamond - metal interface can be expected using heavily N-doped CVD diamond to realise the observed electron emission from the CB. It should also be noted that high resistivity of the heavily N-doped CVD diamond is essential to induce a large voltage drop in the diamond bulk. If the resistivity of the diamond is low, voltage drop in vacuum becomes dominant. In that case, energy barrier at diamond - metal interface cannot be narrowed, and electron injection would not occur.

## Conclusion

As a conclusion, combined XPS/UPS/FES measurements were performed on heavily N-doped polycrystalline CVD diamonds in order to clarify the mechanism of its extremely low-threshold electron emission reported in 1996. Our results clearly prove that the electrons are emitted from the CB of the diamond, utilising NEA. The results can be well explained by the MIV-type electron emission, which the voltage applied between an anode and a cathode drops in diamond bulk, enhancing electron injection from the back contact metal to the CB of diamond due to quantum tunnelling. Moreover, the results imply that this type of electron emission can be achieved using a material, which possesses high resistivity and NEA surface. The low-threshold voltage, extremely narrow energy width of ~0.3 eV, as well as the stable emission current suggests that this type of electron emission can be advantageous in a development of the next generation vacuum nano-electronic devices with long lifetime and high energy resolution.

## Acknowledgements

The authors acknowledge Dr. Shuichi Ogawa of Tohoku University for the experimental supports on the UPS measurement. The authors also acknowledge Dr. Bradford B. Pate of Naval Research Laboratory for the fruitful discussion on UPS and combined UPS/FES of diamond from initial stage of this work. The present work was financially supported by Academic Frontier Project, Support Program for Private Universities (S08 01012) and Grants-in-Aid for Scientific Research (21560053), from the Ministry of Education, Culture, Sports, Science & Technology, Japan.

**Figure captions**

Figure 1
Schematic diagram of combined XPS/UPS/FES system.

Figure 2
Combined UPS/FES spectra of (a) natural type IIb and (b) lightly N-doped CVD diamond. Insets show the FES spectra of each diamond.

Figure 3
Combined UPS/FES spectra of heavily N-doped CVD diamond.

Figure4

Emission current dependence of FE peak energy for heavily N-doped CVD diamond, lightly N-doped CVD diamond, and natural IIb diamond. Inset shows an enlarged plot for the emission from site 2 of the heavily doped diamond.

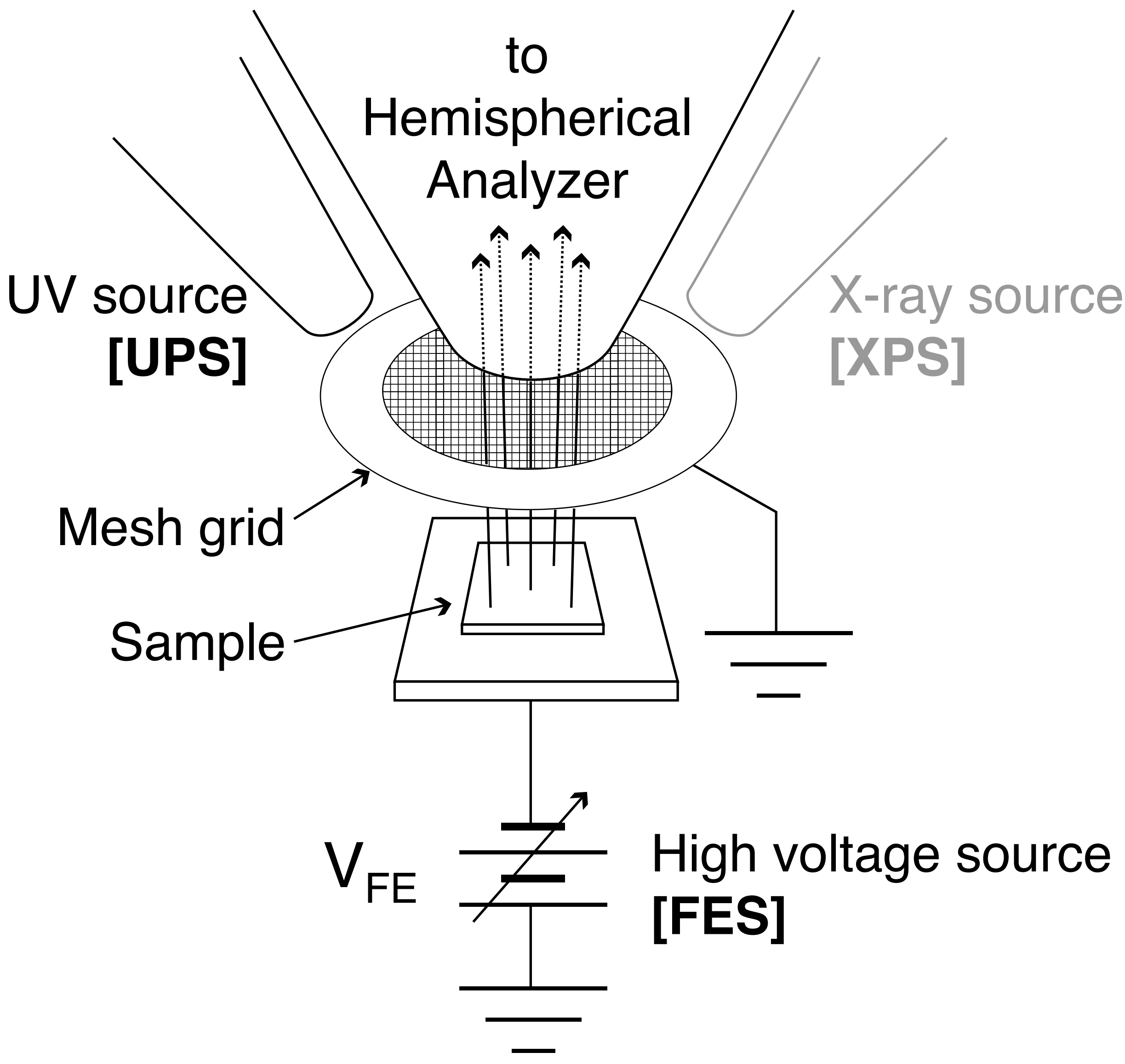

**Figure 1 H.Yamaguchi *et al*.**

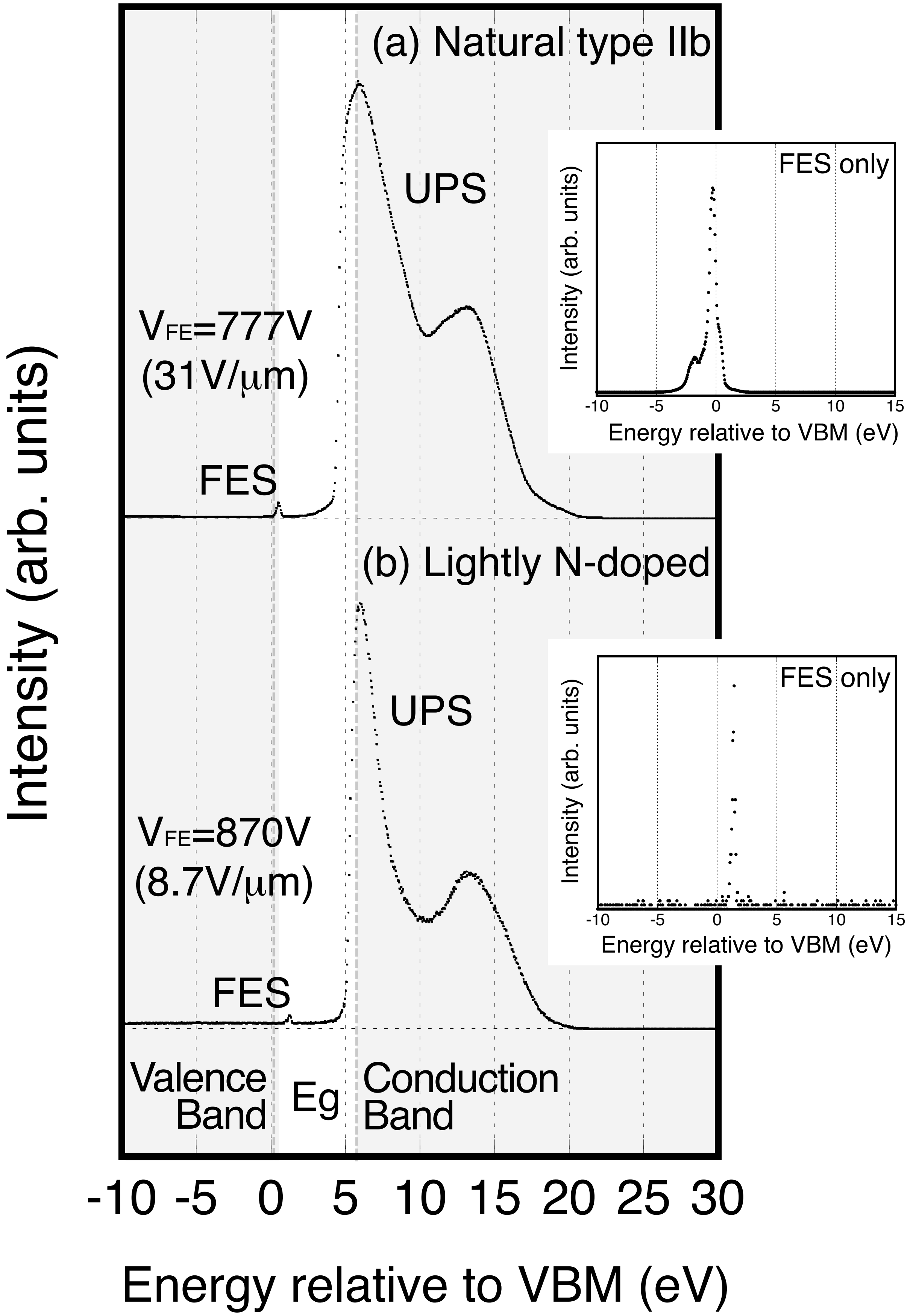

**Figure 2 H.Yamaguchi *et al*.**

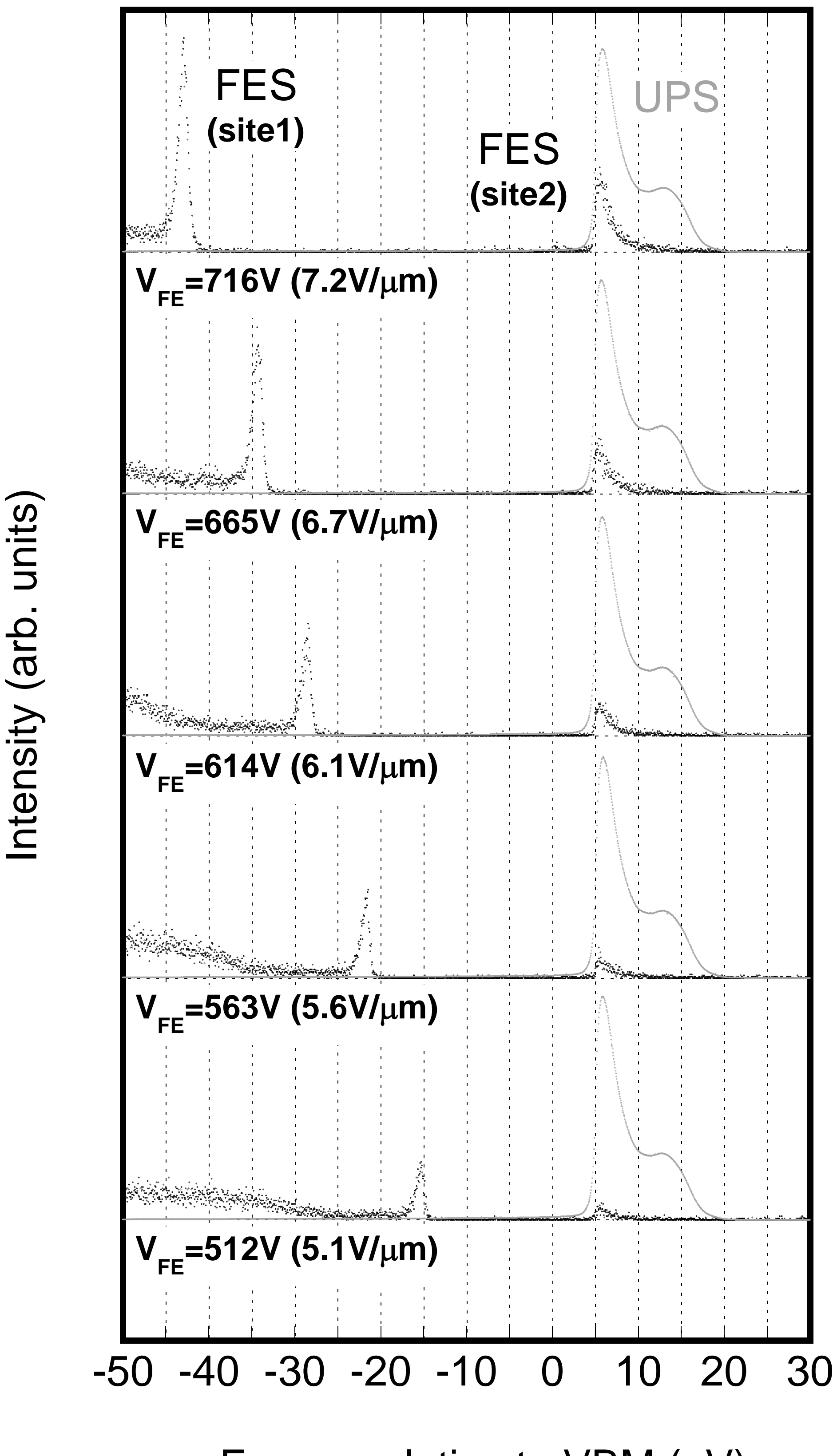

**Figure 3 H.Yamaguchi *et al.***

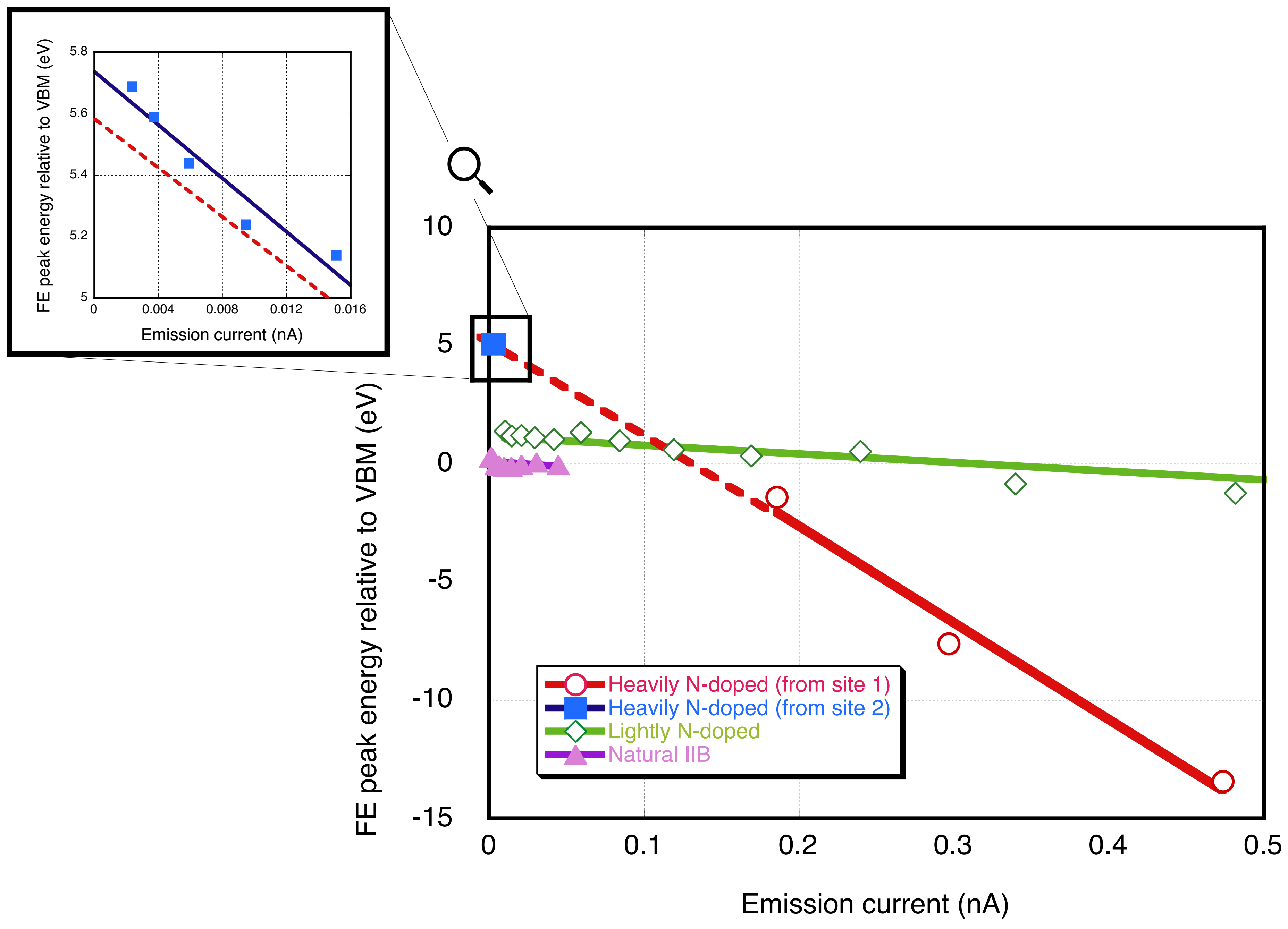

**Figure 4 H. Yamaguchi *et al.***